%% file: mano19.tex
\documentclass[letter,twocolumn]{jpsj3}
\usepackage{txfonts}
\usepackage{color}
\usepackage{bm}
\usepackage{url}
\usepackage{braket}

\newcommand{\newblock{ }}

\title{Application of Convolutional Neural Network to Quantum Percolation in Topological Insulators}

\author{Tomohiro Mano and Tomi Ohtsuki\thanks{ohtsuki@sophia.ac.jp}}
\inst{
Physics Division, Sophia University, Chiyoda-ku, Tokyo 102-8554, Japan} %

\abst{
Quantum material phases such as the Anderson insulator, diffusive metal, and Weyl/Dirac semimetal
as well as topological insulators show specific wave functions both in real and Fourier spaces.
These features are well captured by convolutional neural networks, and the phase diagrams
have been obtained, where standard methods are not applicable.  One of these examples
is the cases of random lattices such as quantum percolation.  Here, we study
the topological insulators with random vacancies, namely, the quantum percolation in
topological insulators, by analyzing the wave functions via a convolutional neural network.
The vacancies in topological insulators are especially interesting since peculiar bound
states are formed around the vacancies.  We show that only a few percent of vacancies
are required for a topological phase transition.  The results are confirmed by independent
calculations of localization length, density of states, and wave packet dynamics.
}

\begin{document}
\maketitle

\input{mano19Content.tex}

\newpage
\bibliography{manoJPSJ19}

\end{document}

%% file: mano19Content.tex

{\it Introduction}--
Three-dimensional (3D) topological insulators (TI)\cite{Hasan10,Qi11,Ando13}
have been attracting considerable research attention.
One of the many interesting features of these materials is the appearance of surface states.
These states are protected topologically, namely, they appear because the topology of
quantum states in the bulk of the material and that of the vacuum are different.\cite{Moore10}

A similar situation is realized when there are spherical or line vacancies in TIs.\cite{Shan11,Imura13} 
In this case, instead of surface states, bound states emerge around vacancies.
The question addressed in this paper is how the bound states are connected when
we have many lattice vacancies, and how the system changes with the increase in the density
of vacancies.

The problem is related to quantum percolation,\cite{Kirkpatrick72,Sur76,Schubert05,StaufferBook,Ujfalusi15}
where the wave functions on the
random lattice begin to be connected.\cite{Makiuchi18} 
The main topic studied here
is the change in the topological nature of the material by the quantum percolation of bound states.

Electron states on random lattice systems are difficult to study, because
 the conventional methods of using the transfer
matrix are not applicable.
The scaling analyses of the energy level statistics\cite{Shklovskii93} are also difficult, if not impossible,\cite{Kaneko99}
owing to a spiky density of states (DoS).\cite{Ujfalusi15}
One way to avoid the difficulty of the transfer matrix is to introduce small but finite transfers in the disconnected bonds,
which is applied to the quantum Hall effect.\cite{Gruzberg17}
Another approach to analyze quantum phase transitions in random lattice TI systems\cite{Chu12} is to calculate
Chern and Z$_2$ numbers.\cite{Agarwala17,Sahlberg19}

Our approach here is to use neural networks.\cite{Tomoki16,Tomoki17,Broecker16,Carrasquilla17,Zhang16,Zhang17b,Yoshioka18,Nieuwenburg17,Zhang18,Araki19,Carvalho18}
 We have shown in refs.~\cite{Mano17,Ohtsuki19} that a method based on a convolutional neural network (CNN)
is free from the above difficulties and works well in determining the phase diagrams of quantum percolation.
In contrast to the study of the DoS where all the information of the energy spectrum is obtained, 
our approach here is to focus on the eigenstates closest to $E=0$, which makes it
 easier and more efficient to draw a global phase diagram.
In this paper, we adopt the idea of using a CNN to draw the phase diagram of percolative 3D topological insulators (3DTIs) in site occupation probability 
vs gap parameter space (see Fig.~\ref{fig:PhaseDiagram}).
The results are confirmed by the calculation of the quasi-one-dimensional (Q1D) localization
length via the iterative Green's function method,\cite{Ando85}  the DoS via the
 kernel polynomial method (KPM),\cite{Weisse06}
as well as wave function dynamics based on the equation
of motion method.\cite{Kosloff94,Ohtsuki97}

{\it Models and Methods}--
We consider the following Wilson--Dirac-type tight-binding Hamiltonian on a cubic lattice\cite{Liu:3DTI,RyuNomura:3DTI},
   \begin{align} \label{eqn:H}
      H = & \sum_{\bm{x}} \sum_{\mu=x,y,z}  V_{{\bm{x}}+{\bf e}_\mu, {\bm{x}}} \left[\frac{{\rm i}t}{2} |{\bm{x}}+{\bf e}_\mu \rangle\alpha_{\mu}\langle\bm{x}|
                                         -\frac{m_2}{2}  |{\bm{x}}+{\bf e}_\mu \rangle \, \beta \langle\bm{x}| + \rm{H.c.}\right]  \nonumber \\
            & + (m_0+3m_2)\sum_{\bm{x}} |\bm{x}\rangle \, \beta \langle\bm{x}|, 
   \end{align}
where $|\bm{x}\rangle$ denotes a four-component state on a site $\bm{x}=(x,y,z)$,
and ${\bf e}_{\mu}$ is a unit vector in the $\mu$-direction. $\alpha_{\mu}$ and $\beta$ are gamma matrices,
   \begin{align} \label{eqn:gammaMat}
      \alpha_{\mu} =\tau_x\otimes\sigma_\mu= \begin{pmatrix}
                         0     & \sigma_\mu \\
                      \sigma_\mu &    0
                   \end{pmatrix}, \ 
      \beta =\tau_z\otimes 1_2= \begin{pmatrix}
                      1_{2} & 0 \\
                      0 & -1_{2}
                   \end{pmatrix}, 
   \end{align}
where $\sigma_{\mu}$ and $\tau_\mu$ are Pauli matrices.
$m_0$ is the mass parameter, and $m_2$ and $t$ are hopping parameters.
In the rest of this paper, we take $m_2=1$ as the energy unit and set $t=2$.
The parameter $V_{{\bm{x}}+{\bf e}_\mu, {\bm{x}}}$ is defined as
\begin{equation}
\label{eq:transfer}
V_{{\bm{x}}+{\bf e}_\mu, {\bm{x}}}=\left\{
\begin{array}{ll}
   1   &  \left({\rm for\;connected\;bond}\right) \,,\\
   0   &  \left({\rm for\;disconnected\;bond}\right)\,.
\end{array}
\right.
\end{equation}
Namely, $V_{{\bm{x}}+{\bf e}_\mu, {\bm{x}}}=1$ if and only if the nearest-neighbor sites $\bm{x}$ and ${\bm{x}}+{\bf e}_\mu$ are connected.
We assumed that a site is randomly occupied with probability $P$ and empty with probability $1-P$, so that the Hamiltonian describes the site percolation model in 3DTIs.
When all the sites are occupied, $P=1$, we can analytically determine the phases\cite{Imura12,Kobayashi15}:
the system is in the ordinary insulator (OI) phase for  $m_0>0$, 
the strong topological insulator [STI(000)] phase for $0>m_0>-2$, 
the weak topological insulator [WTI(111)] phase  for $-2>m_0>-4$,
and  the STI(111) phase  for $-4>m_0>-6$.  The indices such as (000) and (111)
are the weak indices.\cite{Fu07}

Typical eigenfunctions at $E \simeq 0$ are shown in Fig.~1.
We fixed $m_0=-2.5$  so that the system is in the WTI(111) phase in the clean limit, $P=1$.
The fixed boundary condition (FBC) is imposed in the $z$ direction, while periodic boundary conditions (PBCs)
are imposed in $x$ and $y$ directions.
For $P=0.99$, the gapless surface state appears in the $x-y$ plane [Fig. 1(a)] as in the clean limit.
When we reduced the site occupancy to $P=0.97$, the surface states disappear,
and the wave function is extended over the whole system [Fig. 1(b)].
Note that the surface state is destroyed by only 3\% site vacancies despite the robustness of topological materials against disorder.
When we further reduced the site occupancy  to $P=0.90$, the surface state reappears [Fig. 1(c)].

\begin{figure}[htb]
  \begin{center}
\includegraphics[angle=0,width=1.0\linewidth]{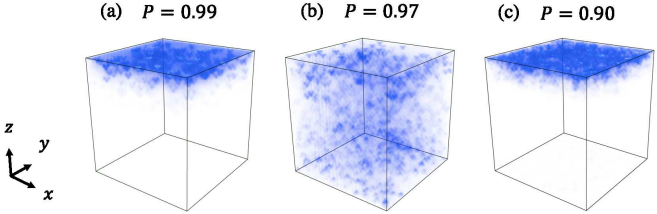}
 \caption{(Color) Typical eigenfunctions of quantum percolation in 3DTI at $m_0=-2.5$ with different site occupancies.
 The gapless surface state appears at $P=0.99$ (a) and $P=0.90$ (c), but is
 destroyed at $P=0.97$ (b).
 }
\vspace{-0.5cm}
\label{fig:WF}
\end{center}
\end{figure}

We first derive a rough phase diagram of quantum percolation in 3DTI.
We adopt  supervised learning with the CNN, which is
an efficient machine learning approach to draw  phase diagrams of random electron systems.\cite{Tomoki16,Tomoki17,Mano17,Ohtsuki19}

In the supervised training, we need a correctly labeled data set (training data) in advance to optimize the weight parameters of the CNN.
Without prior knowledge of quantum percolation in 3DTI, however, preparing enough training data for
each phase is difficult.
We therefore use the weight parameters of the CNN that has learned the features of $k$-space eigenfunctions of the 3DTI with random on-site potential,\cite{Ohtsuki19} where the lattice is regular ($V_{{\bm{x}}+{\bf e}_\mu, {\bm{x}}}\equiv 1$),
but the random on-site potential 
\begin{equation}\sum_{\bm{x}} |\bm{x}\rangle \, V(\bm{x})\,I_4 \langle\bm{x}|\,,
\label{eq:onsite}
\end{equation}
is added with $V(\bm{x})$ random potentials.
Because of the generalization capability of the CNN, we expect it to determine the correct  phases for the unlabeled data set (the wave functions of the 3DTI with vacancies).\cite{Mano17,Ohtsuki19}
An analogy of this approach is that the CNN trained for the Anderson transition with random on-site potential can correctly identify quantum phases in quantum percolation\cite{Mano17}.

To train the CNN, we need to prepare the $k$-space eigenfunctions with varying $m_0$
and random potential.\cite{Tomoki17,Ohtsuki19}
We consider the $24\times 24\times 24$ lattice with the FBC in the
 $z$ direction and the PBC in the other directions, 
diagonalize the Hamiltonian in real space, calculate the eigenfunctions around $E=0$ using the sparse matrix diagonalization Intel MKL/FEAST\cite{Polizzi09},
and obtain the $k$-space eigenfunctions through discrete Fourier transform.
Random numbers are generated by the Mersenne Twister algorithm.\cite{Matsumoto98}


We also study the Q1D localization length for the detailed analysis.
As mentioned in the introduction, in the quantum percolation, the transfer matrix method, an effective method for the case of a regular
lattice, is not applicable
since a matrix relating one layer to the other has zero determinant owing to disconnected bonds.\cite{Ohtsuki19}
In this paper, we employ the iterative Green's function method\cite{Ando85} and calculate the Q1D localization length in the geometry $L \times L \times L_z, \, L_z \gg L$ with PBCs
 in the transverse ($x$ and $y$) directions. (A similar geometry is employed
in the case of wave packet dynamics simulation.)
For numerical simulation, the system size in the $z$ direction is truncated at $L_z = 100,000$.
We shifted the energy slightly from the band center, $E=0$ to $E=0.001$, to avoid the numerical instability of the inverse matrix calculation.
Note that if all the clusters that include sites of the first layer are disconnected at $z<L_z$,
the Green's function method also breaks down.
In practice, this situation does not occur because the parameter regime studied here is well above the classical percolation threshold $P_c\approx 0.312$\cite{Sur76,Wang13}.

In the case of DoS calculation, cubic systems of size $L=160$ with PBCs
are considered.

{\it Results}--
In Fig.~\ref{fig:PhaseDiagram}, we show the phase diagram of quantum percolation in the
3DTI obtained by the CNN.
The abscissa shows the mass parameter $m_0$, while the ordinate is the site occupancy $P$.
From this figure, we see that as the site occupancy $P$ decreases, namely, the disorder increases, the absolute value of the topological mass $|m_0+3|$ 
effectively increases by the renormalization of $m_0$.
The STI(000)  and the STI(111) phases are therefore in contact with each other around $m_0=-3$, and the diffusive metal (DM) phase appears around there
since topologically different phases cannot be connected continuously.
In the case of random on-site potential, in contrast, 
the renormalized mass $|m_0+3|$ decreases with the increase in disorder, 
resulting in the transitions from the OI phase to the STI phase and from STI to WTI (so-called topological Anderson insulator transition\cite{Li09a, Groth09, Guo10,Kobayashi13}).
The difference between our case and that of the random on-site potential indicates that the role
of lattice vacancy disorder is qualitatively different.
We also emphasize that the transition from the WTI(111) phase to the STI phase occurs with less than 5\% lattice vacancies,
despite the common belief that  the topological phase is robust against randomness.

\begin{figure}[htb]
  \begin{center}
\includegraphics[angle=0,width=1.0\linewidth]{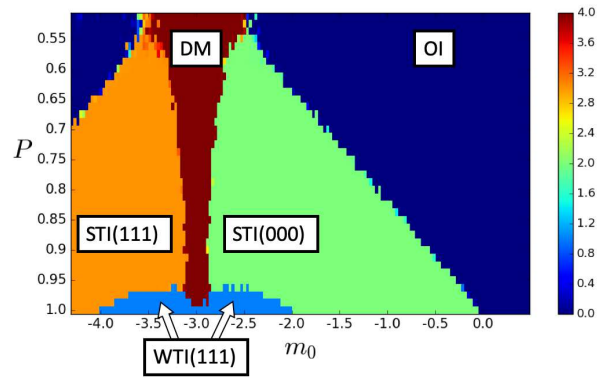}
 \caption{(Color) Phase diagram of quantum percolation in 3D topological insulator.
The CNN outputs the confidence values
$P_{\rm OI}, P_{\rm W111}, P_{\rm S000}, P_{\rm S111}$, and $P_{\rm DM}$
that the wave function belongs to OI, WTI(111), STI(000), STI(111), and DM, respectively,
and the intensity $0\times P_{\rm OI}+1\times P_{\rm W111}+2 \times P_{\rm S000} + 3\times P_{\rm S111} + 4\times P_{\rm DM}$ is plotted.
Sample-to-sample fluctuations are confirmed to be small, so
the sample average is not taken.
 }
\vspace{-0.8cm}
\label{fig:PhaseDiagram}
\end{center}
\end{figure}

 \begin{figure}[htb]
  \begin{center}
\includegraphics[angle=0,width=0.85\linewidth]{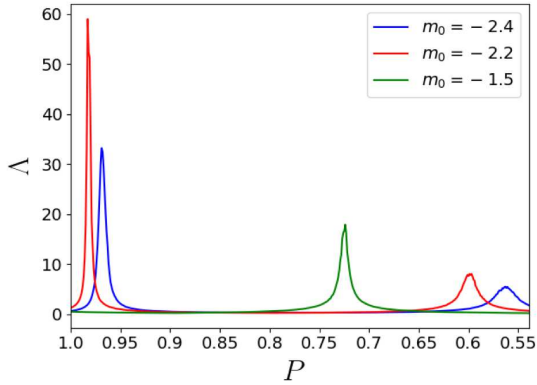}
 \caption{(Color) Normalized localization length $\Lambda$  as a function of $P$ for $m_0=-1.5, -2.2$, and $-2.4$.
The size of the cross section is $L=8$. 
The peak of the localization length indicates the phase boundary where the states
are (semi)metallic and show a larger localization length.
We slightly shift the energy to $E=0.001$ from $E=0$.
We have confirmed that the peak position is insensitive
to the small shift by examining the case for $E=0.0001$.
 }
\vspace{-0.5cm}
\label{fig:LocalizationLength}
\end{center}
\end{figure}
 
 \begin{figure*}[htb]
  \begin{center}
\includegraphics[angle=0,width=0.8\textwidth]{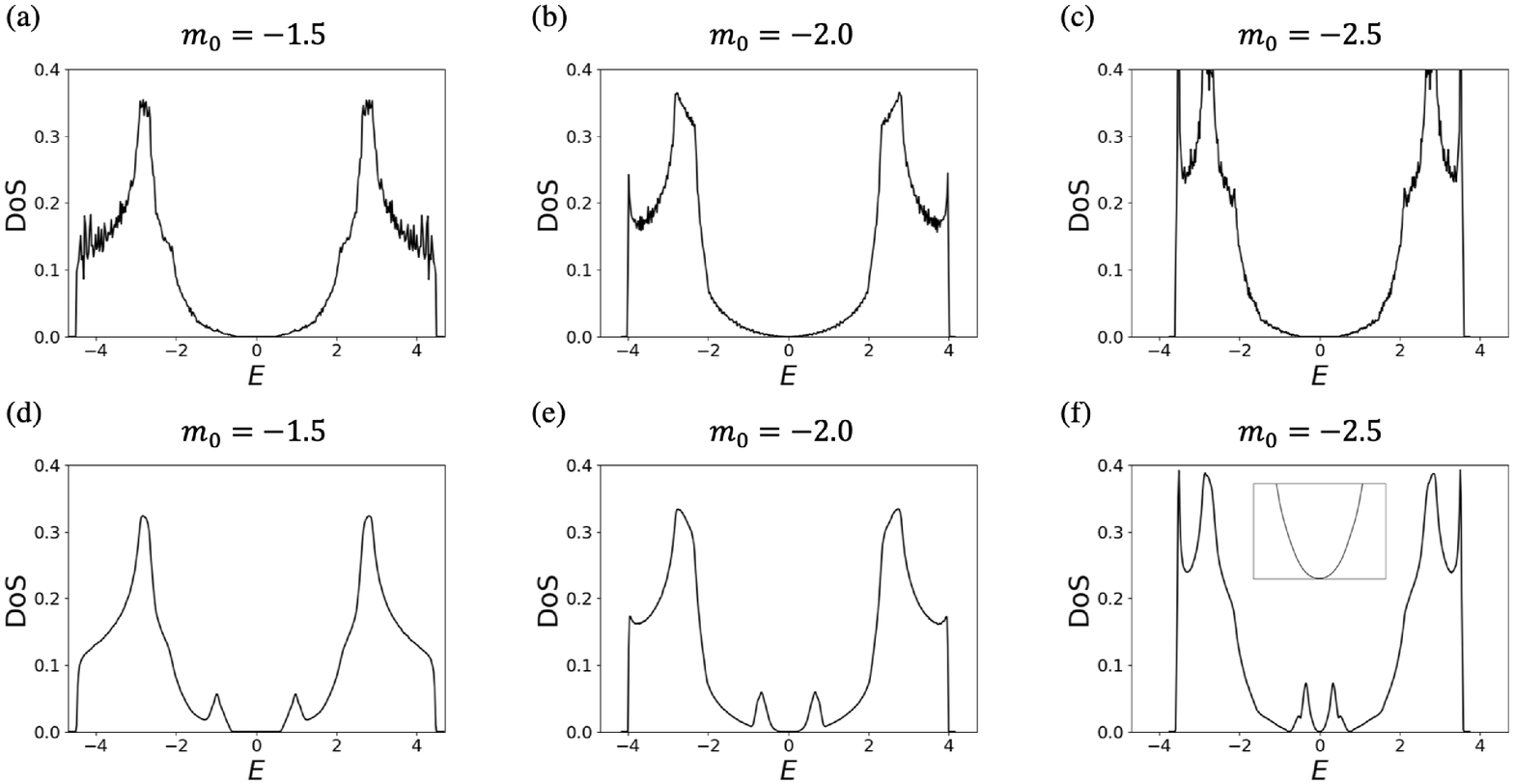}
 \caption{DoSs for $P=1$ [(a)-(c)] and  $P=0.96$ [(d)-(f)]
 calculated by the KPM with the system size $160 \times 160 \times 160$.
 In the presence of lattice vacancies (d)-(f), the minibands formed by the bound states around vacancies appear
 inside the bulk gap.
 At the WTI(111)-STI(000) phase boundary (f), quadratic energy dependence is observed in the vicinity of $E=0$ (see inset). }
 \vspace{-0.5cm}
\label{fig:DoS}
\end{center}
\end{figure*} 

Since the above phase diagram shows unexpected features, 
we verify the phase boundaries by calculating the Q1D localization length $\xi_\mathrm{Q1D}$ via the iterative Green's function method.\cite{Ando85}
Figure \ref{fig:LocalizationLength} shows the normalized localization length\cite{MacKinnon81,Pichard81,Kramer93}
$\Lambda=\xi_\mathrm{Q1D}/L$ as a function of $P$ for $m_0=-1.5, -2.2$, and $-2.4$.
We see that for each $m_0$, the peak of the localization length is located at the phase boundary in Fig.~\ref{fig:PhaseDiagram}.
At $m_0=-2.4$, for example, the localization length exhibits two peaks around $P=0.97$ and $P=0.56$, corresponding to the phase boundary of WTI(111)-STI(000) and that of
 STI(000)-OI, respectively (Fig.~\ref{fig:PhaseDiagram}).
 This increase in Q1D localization length at the phase boundary
 is consistent with the well-known fact that Dirac semimetals (DSMs) continue to exist at the topological phase boundary
 even in disordered systems.\cite{Kobayashi14,Syzranov15}

To further understand the nature of the topological phase transition due to vacancies,
we also study the DoS using the KPM\cite{Weisse06}.
The DoSs in the clean limit, $P=1$, at $m_0=-1.5, -2.0,$ and $-2.5$ are shown in
Figs.~\ref{fig:DoS}(a)-\ref{fig:DoS}(c), respectively.
The case $m_0=-2.0$ (b) corresponds to the phase boundary between the STI(000) and the WTI(111) phases,
on which the system is a DSM.  The linear energy dispersion around $E=0$ leads to a
 parabolic DoS, as is clearly seen in Fig.~\ref{fig:DoS}(b).

The DoSs with 4\% lattice vacancies, $P=0.96$, with the same $m_0$'s are shown in 
Figs.~\ref{fig:DoS}(d)-\ref{fig:DoS}(f).
When lattice vacancies are present, the bound states appear around them,
and these bound states form minibands inside the original band gap.
In Fig.~\ref{fig:DoS}(f), the miniband shows a parabolic DoS in the vicinity of $E=0$ (see the inset),
which is consistent with the phase diagram: the parabolic DoS appears for
 the parameter set $(m_0,P)=(-2.5,0.96)$, which is on the phase boundary in Fig.~\ref{fig:PhaseDiagram}.
In the following, we show that this parabolic DoS in the miniband comes from DSM states
on the phase boundary.

 \begin{figure}[htb]
  \begin{center}
\includegraphics[angle=0,width=1.0\linewidth]{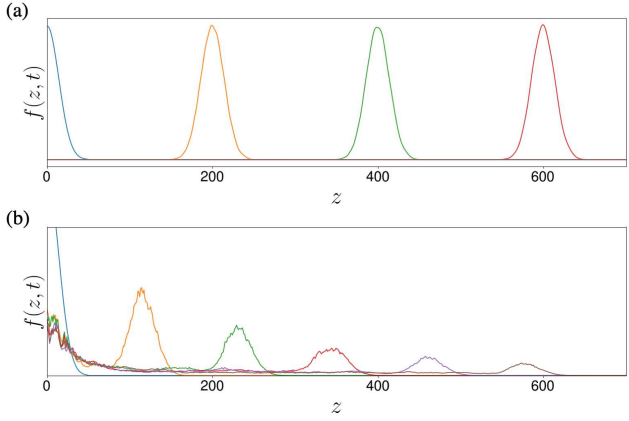}
 \caption{(Color) Time evolution  of wave packet in the Q1D geometry with cross section of $20 \times 20$. 
 The parameter sets are (a) $(m_0,P)=(-2.0,1.0)$ and (b) $(m_0,P)=(-2.3, 0.975)$.
 We plot  the probability density $f(z,t)=\int dx dy\, |\psi (x,y,z,t)|^2$ of a wave packet $ \psi (x,y,z,t)$.
The snapshots in (a) are for $t=0, 100, 200, 300 (\hbar/m_2)$, while
those in (b) are for $t=0, 100, 200, 300, 400, 500 (\hbar/m_2)$.
 The initial Gaussian wave packet is prepared at $z=0$.
  Averaging over five samples has been performed.
Amplitudes near $z=0$ in (b) are the remains of ballistic wave packets.
 }
\label{fig:EQM}
\end{center}
\end{figure}

A DSM is characterized by its ballistic transport.
To confirm this, we employ the Chebyshev polynomial expansion for the time-evolution operator\cite{Kosloff94} $U(\Delta t)=\exp (-iH\Delta t)$,
and study the wave packet dynamics in the Q1D geometry.
The results for two parameter sets corresponding to the phase boundaries, (a) $(m_0,P)=(-2.0,1.0)$ and (b) $(m_0,P)=(-2.3, 0.975)$, are shown in Fig.~\ref{fig:EQM}.
The ballistic transport in the absence of disorder [Fig.~\ref{fig:EQM}(a)] survives
even in the presence of randomness [Fig.~\ref{fig:EQM}(b)].
The decay of the wave packet in Fig.~\ref{fig:EQM}(b) is because
 part of the wave packet is trapped by the bound states around the vacancies during the transport process\cite{KobayashiEQM,KobayashiPrivateComm}.
From the time dependence of the peak positions, we evaluate the speed $v$ of the ballistic transport and obtain (a) $v_{\rm a}=t a /\hbar=2\times m_2 a/\hbar$ ($a$ being the lattice constant) and (b) $v_{\rm b}=1.15\times m_2 a/\hbar$.
The ratio $\alpha$ of the renormalized velocity to the bare one,
 $\alpha = v_{\rm b}/v_{\rm a}$, is estimated to be  0.575.
We also estimate the curvatures of the DoSs around $E=0$
by fitting the DoS with a quadratic polynomial in the vicinity of $E=0$,
 and obtain (a) $\rho_{\rm a}\approx 0.0103\,E^2$ and (b) $\rho_{\rm b}\approx 0.0536\,E^2$.
From $\rho \propto E^2/v^3$, we estimate $\alpha'=(\rho_{\rm a}/\rho_{\rm b})^{1/3}=0.577$,
 in good agreement with the estimate of the ballistic velocity ratio $\alpha$.
We therefore conclude that the parabolic DoSs inside the minibands [e.g., Fig.~\ref{fig:DoS}(f)] are formed by a DSM.

{\it Summary and concluding remarks}--
In this paper, we have shown that a small amount of vacancies induces
a topological phase transition.  For example, a weak topological insulator
undergoes topological phase transition and becomes a strong topological insulator
with only a few percent of vacancies.
A few percent of vacancies, at first sight, may sound very small.
Around a vacancy, however, a bound state on the order of 10 sites is formed.
Such bound states therefore fill a significant amount of sites even for a few percent of vacancies,
resulting in metallic states that spread over the system [see Fig.~\ref{fig:WF}(b)].
We emphasize that the vacancy-driven topological transition is in sharp contrast to the case of on-site random potential, where the topological
phase transition does not occur as long as the bulk band gap remains.  
Note also that
contrary to the present problem, in the case of on-site potential, the strong-to-weak topological insulator and
the ordinary-to-strong topological insulator transitions take place
with the increase in randomness.

The Hamiltonian considered here preserves the particle-hole symmetry even
in the presence of vacancies, i.e., $ C H C^{-1} =-H^T\,,\, C=\tau_y\otimes\sigma_y\,,\,
C=C^T$, and belongs to the class DIII.\cite{Zirnbauer96,Altland97,Schnyder08}
Furthermore, we have an additional sublattice symmetry for $m_0=-3m_2$,
where the last term in Eq.~(\ref{eqn:H}) vanishes.
The latter symmetry is rather artificial, since this term may fluctuate and deviate from 0
in actual materials, and some features of the phase diagram such as
symmetry around $m_0=-3m_2$ will not be observed in real materials.
The sensitivity of topological phases against vacancies, however, remains
even in the presence of site-to-site fluctuation of $3 m_2$.
We have confirmed that changing the last term in Eq.~(\ref{eqn:H}),
$ (m_0+3m_2)\sum_{\bm{x}} |\bm{x}\rangle \, \beta \langle\bm{x}|$,
to $ \sum_{\bm{x}} (m_0+m(\bm{x})/2)|\bm{x}\rangle \, \beta \langle\bm{x}|$
with $m(\bm{x})/m_2$ given by the number of bonds connected to the site $\bm{x}$
breaks the symmetry of the phase diagram around $m_0+3m_2$,
but the overall features of the phase diagram remain unchanged.
The addition of on-site randomness, Eq.~(\ref{eq:onsite}), with $V(\bm{x})$
independently and uniformly distributed in $[-W/2,W/2]$,
derives the system to the Wigner--Dyson symplectic class AII,\cite{Wigner51,Dyson61,Dyson62,Altland97,Schnyder08} but as long as
$W$ is small, say $W=0.5$, the  features of
the phase boundaries remain almost unchanged.

\bigskip
\noindent
{\bf Acknowledgement}
The authors would like to thank Dr. Koji Kobayashi for useful discussions. This work was partly supported by JSPS KAKENHI Grant Nos. JP15H03700, JP17K18763, 16H06345, and 19H00658.
